\documentclass[aps,11pt]{revtex4}
\usepackage{graphicx}

\begin{document}
\title{Dissipationless Flow and Sharp Threshold of a Polariton Condensate with Long Lifetime}
\author{Bryan Nelsen, Gangqiang Liu, Mark Steger, and David W. Snoke\footnote{email address: snoke@pitt.edu}}
\affiliation{Department of Physics and Astronomy,
University of Pittsburgh, Pittsburgh, PA 15260, USA}
\author{Ryan Balili\footnote{present address: Cavendish Laboratory, Cambridge University, Cambridge CB3 0HE, UK}}
\affiliation{Department of Physics, MSU-Iligan Institute of Technology, Iligan 9200, Philippines}
\author{Ken West and Loren Pfeiffer}
\affiliation{Department of Electrical Engineering, 
Princeton University, Princeton, NJ 08544, USA}

\abstract{
We report new results of Bose-Einstein condensation of polaritons in specially designed microcavities with very high quality factor, on the order of $10^6$, giving the polariton lifetimes of the order of 100 ps. When the polaritons are created with an incoherent pump, a dissipationless, coherent flow of the polaritons occurs over hundreds of microns, which increases as density increases. At high density, this flow is suddenly stopped, and the gas becomes trapped in a local potential minimum, with strong coherence.}
}

\maketitle

Previous work by many experimental groups has shown that polaritons in microcavities at high density and low temperature act in many ways the same as Bose-Einstein condensation (BEC) of cold atoms, with numerous resulting effects such as quantized vortices \cite{devvort}, Josephson oscillations between two condensates with time-dependent period that depends on the particle density \cite{devJJ}, bimodal momentum distribution \cite{dev,science07}, indications of a Bogoliubov linear excitation spectrum \cite{yamabog,devbogo,assman} and many other fascinating effects (for reviews see Refs.~\cite{pt,yamareview,kavbook}). Essentially, what is done in these experiments is that photons, which are normally massless and noninteracting, are given an effective mass by means of a cavity and are given mutual interactions by means of a resonance with an electronic excitation in the medium, namely a semiconductor quantum well exciton. The effective $\chi^{(3)}$ in these systems reaches world-record values, around four orders of magnitude higher than typical nonlinear solid optical media \cite{snokechap}, and this makes the system behave in ways quite different from a standard laser. The mathematical description of the interacting photons, that is, the polaritons, is given by the theory of a weakly interacting Bose gas in two dimensions, with the addition of generation, drag, and decay terms. 

Because of the short lifetime of the polaritons in previous structures, long-range coherent flow has been hard to observe. In one set of experiments \cite{madrid}, a coherent condensate was made by direct resonant pumping over a large area, and a spatially compact pulse in this condensate was observed to propagate over long distances. Since the condensate was created by pumping with a laser with energy resonant with the polariton energy, it could be argued that this effect was simply the propagation of the laser coherence. In another set of experiments \cite{bloch-wire}, polaritons were created with an incoherent pump in quantum wire structures, and showed coherent propagation over distances of about 30 microns. It was not clear in those experiments what role the dimensionality played in the propagation, since transport is strongly altered in one-dimensional structures; the quantum wire system with repulsive polaritons may resemble more closely a Tonks-Girardeau gas \cite{caru}. In those experiments, the coherent state also existed in a multimode state at high density, with several discrete energy levels in the quantum wires. A true Bose condensate is expected to resist multimode behavior (see, e.g, the argument by Nozier\'es in Ref.~\cite{noz}, made more rigorously in Ref.~\cite{comb}) when there are interactions between the particles. The multimode behavior was therefore not strongly different from that of a multimode laser with discrete states due to quantum confinement. Finally, recent experiments \cite{baum-naturephys} have shown stable patterns of polaritons in two-dimensional traps, also over length scales of about 30 $\mu$m, which correspond to the standing waves of a coherent, bounded wave. Strong dissipation was seen in those experiments as the particles lost energy from their initial state  \cite{baum-flower}, as the polaritons fell toward the standing wave states which corresponded to the confined states of the trapped geometry. Like the experiments of Ref.~\cite{bloch-wire}, the coherent modes appeared in a multimode state similar to a multimode laser. 

In this paper, we report experiments with new polariton microcavity structures with extraordinarily long-lived polaritons. With lifetimes of the order of 100 ps, the particles can propagate over long distances, far from the incoherent pump region, and can scatter with each other many times so that many-body effects are important. We see a robust dissipationless and coherent flow over hundreds of microns, well away from the incoherent pump region, and with a strong resistance to multimode behavior, and at high density we see a sudden transition to a trapped state. 

{\bf Experimental Methods}. The polaritons in these structures exist in a two-dimensional plane, defined by the two parallel mirrors which make up the optical cavity and by three parallel sets of four GaAs quantum wells between the mirrors, placed at the three antinodes of the confined optical cavity mode. The only important difference between our new structures and our old structures \cite{science07,2thres} is that the cavity formed by the distributed Bragg reflectors (DBRs) used to make the microcavities now has a quality factor (Q) of around $10^6$, with a designed cavity lifetime of 400 ps. Our previous experiments used a structure which was identical in design to this one, but with half as many layers in the DBR mirrors. The designed cavity lifetime for that structure was 1.5 ps, with a Q of 4800.  The high Q calculated for our new structure is confirmed experimentally by the intrinsic line width of the lower polariton reflectivity line, which is narrower than our instrumental resolution of 0.05 nm.  A direct measurement of the lifetime is not simple in microcavity polariton systems, because incoherent generation of the polaritons leads to the creation of other carriers which can feed into the polariton population, and direct excitation of the polaritons involves a coherent ring-up as well as ring-down of the polariton resonance.  There is also some confusion in reporting polariton lifetimes, because the polariton population lifetime can be much longer than the cavity photon lifetime. The polariton lifetime is a function of the photon fraction, which decreases to zero as the polariton momentum increases; all structures used for microcavity polariton experiments prior to the one we report here have a cavity lifetime of less than ten picoseconds, although the polariton population lifetime could be 20-30 ps. A basic consistency of many measurements, some of which are discussed here and some of which will be reported elsewhere \cite{steger}, gives us a consistent picture of the cavity lifetime in these new structures of the order of 100 ps, and a polariton population lifetime even longer than that. 

The cavity width in our structures varies continuously as a function of the distance from the center of the wafer. Consequently, the cavity photon energy on one end of the structure is below the exciton energy in the GaAs quantum wells, while on the other end of the structure, the photon energy is above the exciton energy. When the photon energy is below the exciton energy, the lower polariton state is mostly photon-like; this is called negative detuning, with the detuning parameter $\delta = E_{\rm photon}-E_{\rm exciton}$. In the reverse situation, with $\delta > 0$, the polaritons are mostly exciton-like. Exactly at the resonance point on the sample where the photon and exciton energies are equal, the polaritons are 50-50 quantum superpositions of a photon and an exciton, with an energy splitting between two polariton states known as the Rabi splitting, which in this case is approximately 13 meV. In our samples this Rabi splitting is large compared to the intrinsic line widths of the photon and exciton modes; this is known as the strong coupling limit. 

The interaction of the polaritons with each other and with the phonons in the lattice comes entirely through their excitonic part. Therefore, by moving to a region of the sample with negative detuning, we reduce the interactions. There is a tradeoff in what detuning to choose. When the particles are more excitonic, they will thermalize much better to the lattice. Polaritons reaching nearly the lattice temperature when they are excitonic has already been reported for short-lifetime polaritons \cite{dengtherm}; we have also seen this behavior at high polariton density in these long-lifetime structures, as we will report elsewhere \cite{wentherm}. In this strongly excitonic regime, however, the mean free path, and therefore the propagation distance, of the polaritons will also be shorter. For the experiments discussed here, we have chosen a region on the photonic side of the sample where the polaritons have relatively weak but nonzero interactions with the lattice and with each other. This allows them to have long distance motion, while still having enough interaction between themselves for many-body effects to be important. 

In all of the experiments discussed here, the excitation laser photon energy is well above the polariton energy, with wavelength 707 nm, at the third minimum of the reflectivity above the microcavity stop band. Since the excess energy is large, many phonons were emitted by the excited carriers before they turned into polaritons; as a result, the coherence of the pump laser is lost to the incoherent phonon bath and not transmitted to the polaritons. A continuous-wave, intensity-stabilized diode laser was used to minimize the role of power fluctuations \cite{skol}. The bath temperature in which the sample was held was approximately 7 K. A Princeton CCD camera on a 0.3 m spectrometer was used for time-integrated spectral images of the light emission from the polaritons.  

{\bf Transition to dissipationless flow}. One of the beautiful aspects of polariton experiments is that the momentum distribution of the particles can be recorded directly by angle-resolving the far-field photon emission from the polaritons (see the supplementary information for details), because the in-plane $k$-vector of a polariton, $k_{\|}$, has a one-to-one correspondence with the external angle of photon emission. Figure~\ref{fig.kspace} shows the $k$-space resolved data under identical conditions except for the incoherent pump power, which is increased from (a) to (c).  The polaritons were created at a spot on the sample on the photonic side of the resonance (detuning $\delta = -3.2$ meV). As discussed above, the shift of the lower polariton energy with detuning due to the gradient in the cavity width gives a spatial gradient of the potential energy felt by the polaritons, which corresponds to a force on them, $F= -\nabla U(x)$. 

The white parabola in these images corresponds to the single-particle energy dispersion $E(k)$ for polaritons at low density at the point of creation. Both Fig.~\ref{fig.kspace}(a) and (b) are ``smeared'' leftward, away from this parabola, because these images spatially integrate the emission from the polaritons as they move in their 2D plane. The polaritons in our structure have such long lifetime that they can move significantly away from the point of creation, to locations in the sample where the potential energy is quite different. As a result, one can think of the $k$-space images of Fig.~\ref{fig.kspace} as ``tomographic'' images in which we can watch the evolution of the momentum of the polaritons as they move spatially \cite{wertz2}. The images are smeared to the left in $k$-space because the potential gradient they feel, discussed above, gives them a leftward force, and a force is equal to $F = \hbar (\partial k/\partial t)$.

The transition from the behavior of Fig.~\ref{fig.kspace}(a) to Fig.~\ref{fig.kspace}(b) is striking. In (a), the polaritons are moving essentially ballistically, with little scattering with each other, due to their low density, and very little scattering with phonons, due to their very light mass \cite{hartwell}. The dynamics in this semiclassical regime are quite interesting in their own right, giving constraints for the scattering processes of the polaritons and their lifetime. This analysis will be presented elsewhere. As the density of the polaritons has increased, there is a gradual transition to a distribution like that shown in Fig.~\ref{fig.kspace}(b); the energy distribution of the polaritons shifts more and more to lower energy until a large fraction are all at a single energy.  

The blue shift of the polaritons in Fig.~\ref{fig.kspace}(b) relative to the ground state of the single-particle $E(k)$ can be understood by looking at the real-space images of the same data. Fig.~\ref{fig.realspace} shows the real-space data under the nearly the same conditions and the same sequence of powers as in Fig.~\ref{fig.kspace}. The data of Fig.~\ref{fig.realspace}(a) shows hot carrier luminescence which gives the extent of the laser excitation spot. As seen in Figs.~\ref{fig.realspace}(b) and (c), the polaritons move very far away from this spot. 

The spatial motion of the polaritons seen in Fig.~\ref{fig.realspace}(b) is at first surprising: the polaritons move uphill, to higher potential energy, for hundreds of microns; in fact, uphill motion more than 1 mm has been observed in this structure, as discussed in the supplementary material. This uphill motion can be understood if we take into account the fact that the polaritons are initially created in a wide range of $k$-states, as seen in Fig.~\ref{fig.kspace}(a). Some of these are moving downhill in the spatial energy gradient while others are moving uphill. The data of the images in Fig.~\ref{fig.realspace} correspond to collecting only low-momentum, $k_{\|} \simeq 0$ emission (i.e., small angle acceptance, since momentum $k$ maps to angle of emission, as discussed above). Therefore those polaritons which move downhill accelerate to high $k$ and out of the angle of acceptance of our imaging system. Those which move uphill with high momentum are also initially out of our field of view, but these slow down (corresponding to the right side of Fig.~\ref{fig.kspace}(a), which is smeared toward the left) and eventually reach $k_{\|} = 0$ when their energy equals the potential energy of the uphill slope. We can call this the ``turnaround point.'' Some of these polaritons actually continue through $k_{\|}=0$ to negative values of $k_{\|}$, as seen in Fig.~\ref{fig.kspace}(a), which corresponds to turning around and accelerating back in the other direction.

As seen in Fig.~\ref{fig.realspace}(c), when the pump is increased to higher density, the polaritons are launched from a higher potential energy due to the exciton cloud which is also created by the incoherent pump. The excitons, with mass $10^4$ times higher than the polaritons, are essentially a static barrier which does not move, as seen by the polaritons; the properties of this potential can be modeled well as a simple mean-field shift proportional to the local density of the excitons. This effect has been seen before, e.g., in Refs.~\cite{bloch-wire} and \cite{baum-naturephys}. The white dashed line in Fig.~\ref{fig.realspace}(c) indicates the height and spatial extent of this potential, on top of the constant gradient of potential due to the cavity thickness gradient. After they come into existence on top of this exciton cloud, the polaritons are accelerated away from this spot by the steep potential energy gradient on the sides of the exciton cloud.  The ones accelerated uphill appear as a bright spot at $k_{\|}\simeq +1.5 ~\mu$m$^{-1}$ in Fig.~\ref{fig.kspace}(b), while the ones accelerated downhill appear at $k_{\|} \simeq -1.5 ~\mu$m$^{-1}$ in the same figure. This corresponds to the momentum the particles have when they convert all of the potential energy they start with at the top of the exciton cloud into kinetic energy; this scenario has been modeled already for short-lifetime polaritons in Ref.~\onlinecite{wouters}. The polaritons then stay at the same energy while moving. In the image of Fig.~\ref{fig.realspace}(c), the polaritons appear at the point of creation and then disappear as they move to higher $k_{\|}$ both to the left and right, as they are accelerated to high momentum, outside our angular field of view which is restricted to $k_{\|} \simeq 0$. The polaritons traveling uphill eventually hit a turnaround point several hundred microns away, at which point they have slowed to have $k_{\|} \simeq 0$ again and are visible in our imaging system. When light emission from the turnaround point and light emission from the creation spot are overlapped, we see fringes when there is a time delay of 140 ps (the travel time from one spot to the other), indicating that the propagation is coherent. The measured coherence time of this monoenergetic beam is approximately 40 ps; the supplementary material for this paper gives further discussion of the propagation of the coherence, which is maintained over travel distances of more than 200 $\mu$m.  

The behavior in the case of Figs.~\ref{fig.kspace}(b) and \ref{fig.realspace}(c) is quite different from the low density case, however. As seen in both Figs.~\ref{fig.kspace}(b) and \ref{fig.realspace}(c), the transport is monoenergetic, meaning there has been no significant loss of energy of the particles over macroscopic distances. No multimode behavior is ever seen. The single-mode behavior seen here occurs when one state is selected out of the continuum of 2D planar $k$-states, as opposed to discrete trapped states as in Refs.~\cite{bloch-wire} and \cite{baum-naturephys}. 

As discussed above, the monoenergetic, coherently flowing fraction of the polaritons increases as the density increases, until almost the entire polariton distribution is participating in this flow. We interpret this as coming about due to the Bose-Einstein statistics of the polaritons, which cause the polaritons to increasingly accumulate in their ground state at their point of creation as the density is raised, due to the stimulated scattering final-states factor $(1+N_f)$ \cite{hartwell}. There is no other mechanism which can cause this kind of nonlinear change in the polariton energy distribution. Once the polaritons are in their ground state at the point of creation, they stream away from the creation spot, since there are no confining boundaries to their flow.  The polaritons far from the creation spot have much lower density as they stream away. In this regime, their interactions are too weak to redistribute their energies; they simply flow ballistically nearly no scattering for hundreds of microns. Thus, while they may start out as a superfluid at the creation spot, once they are in the ground state, they can simply flow coherently without scattering.   

As discussed above, streaming coherent flow of polaritons has also been reported elsewhere \cite{bloch-wire,baum-flower,wouters,kasp-ring,kasp-thesis} under non-resonant excitation, but with short-lifetime polaritons. In that case there was also an acceleration of a coherent propagagating state away from the potential energy peak created by the exciton cloud, but only over a very short distance near the laser excitation spot; there was also some degree of multimode behavior. In the present experiments, the long lifetime of the polaritons allows them to propagate completely away from the laser generation spot, all the way to the point at which they turn around and come back in the potential gradient created by the cavity. 

{\bf Critical threshold for a trapped state}. As seen in Figs.~\ref{fig.kspace}(c) and \ref{fig.realspace}(d), when the pump power exceeds a critical threshold, the behavior of the system changes dramatically again, with a drop in energy. The polaritons drop to the ground state of the quasi-1D trap created in the nook between the exciton cloud and the sloping potential of the cavity gradient. The polariton gas now has a very tightly compact spatial profile with a radius of approximately 5.5~$\mu$m. Fig.~\ref{fig.thres} shows that the threshold for this behavior is very sharp. No equivalently sharp jump in spectral narrowing has been seen before in experiments with short-lifetime polariton systems.\cite{baum-new} The coherence time of this spot is greater than the limit measurable by our Michelson interferometer, 280 ps.  As discussed below, we believe the actual coherence time is of the order of 1 millisecond.

The exact model for this sudden drop in energy and flow to a trapped state is an open question for theory.  We can make several observations, however. First, as mentioned above, the polaritons have almost no interaction with the lattice phonons \cite{hartwell}, and elastic scattering with impurities does not change their energy. Since the excitons at the creation point are so heavy, scattering with excitons will also be effectively elastic and will not change the polariton energy. On the other hand, the polariton-polariton interaction is not negligible, even away from the high-density point of creation. We can see in the trapped state of Fig.~\ref{fig.realspace}(d) that the interaction energy of the polaritons, which in this case is pure polariton-polariton interaction, since the polaritons have moved away from the laser spot, is approximately 0.2 meV. It may be that at high density, the polaritons returning from the turnaround point (seen in Fig.~\ref{fig.realspace}(c)), collide with the polaritons streaming outward from the creation spot, giving them a new channel for energy redistribution.  fro

The density threshold at which this trapped state occurs is comparable to the density at which equilibrium Bose-Einstein occurs in a trapped geometry, in work to be reported elsewhere \cite{wentherm}. It is therefore reasonable to assume that the sudden transition is to a truly coherent condensate state. 

Another indication of the high coherence of this trapped state is shown in Figure~\ref{fig.vort}, which shows two-dimensional $k$-space images of the polariton emission. The only difference in experimental conditions of the three different images is that they were taken at different times, with 5 ms exposure time while a steady-state laser pumped the system. These images are not taken with interferometry---the condensate itself has fringes which are stable for time periods of around a millisecond. The variations in time are presumably related to small fluctuations of the pump laser intensity. Berloff \cite{berloff} has modeled the fringes as the result of vortices due to the collision of inflowing condensate with a trapped condensate in steady state. As seen in Fig.~\ref{fig.vort}, the coherent polariton wave tunneling passing around the exciton cloud barrier, giving a high-momentum tail, also has fringes, which can be associated with vortex streets \cite{vortex1,vortex2}.

The fact that the energy of this emission is {\em lower} than the energy of the emission at lower pump powers shows that this emission is not standard lasing---standard lasing has been shown to occur in this type of structure, and always occurs at higher energy, at the cavity photon mode \cite{2thres,JAP,ass-2thres}. In the data shown in Figs.~\ref{fig.kspace} and \ref{fig.realspace}, the bare cavity photon energy at the point of creation is at 1.60 eV, well above the relevant polariton energies. Also, the fact that the trapped state is spatially displaced from the location of the laser pump spot by about 16 $\mu$m, and is in a region with almost no pump laser intensity, also indicates that standard pumped lasing is not occurring. 

{\bf Conclusions and Future Directions}. The overall picture which emerges is that there are three regimes of behavior. At low density, the polariton gas is classical and has a Maxwellian energy distribution, with drift due to flow in an energy gradient, as seen in Fig. 1(a). As density is increased, the energy distribution shifts toward greater and greater occupation of the lowest energy state at the point of creation, as seen in Fig.~\ref{fig.thres}, due to the Bose statistics. This nearly monoenergetic, but dephased, population can be called a ``quasicondensate'' \cite{quasi}.  A gradual increase of the occupation of the ground state is predicted by the Bose-Einstein statistics of the particles in two dimensions. At a higher density, there is a very sharp transition to a third regime, which has all the characteristics of a ``true'' condensate in a trap. The sharp transition may be due to a Berezhinski-Kosterlitz-Thouless superfluid state \cite{stringari,BKT2}, or it may be due to the onset of trapping due to the interaction with growing exciton cloud; a trap allows a true condensate in two dimensions \cite{BKT2,baym}. The trapped state is characterized by stable patterns which are consistent with vortex formation. 

Motion and acceleration of microcavity polaritons has been seen before \cite{assmanstep,wedge}, but never over such long distances. This is a quantitative improvement from distances of the order of 10~$\mu$m to distances of up to a millimeter. The long-range propagation allows us to do polariton physics entirely in the absence of any pump; in much theory of polariton condensates up to now, it has been assumed that the polaritons are in constant contact with the exciton ``reservoir'' created by the pump laser, although previous experiments with 20-30 micron propagation \cite{bloch-wire,baum-naturephys} have begun to encourage theoretical calculations without a uniform exciton background.

As reported elsewhere, we are able to make a controlled harmonic potential trap for the polaritons using inhomogenous stress \cite{apl,science07}. We have recently succeeded at making ring condensate, in which a Mexican-hat potential is created using an exciton cloud in the center of a harmonic potential trap. A similar sudden threshold to a trapped state in the ring as is reported in this work. As in the results reported here, this condensate is spatially in a different location from the pump laser. These results will be reported elsewhere \cite{ring}. It is also possible now to use the AC Stark effect to create a potential for the excitons on very short time scales \cite{hayat}.  These new structures with long polariton lifetime open up qualitatively new physics on a macroscopic scale. 

{\bf Acknowledgements}. This material is based upon work supported by the National Science Foundation under grant  DMR-1104383. The work at Princeton was partially funded by the Gordon and Betty Moore Foundation as well as the National Science Foundation MRSEC Program through the Princeton Center for Complex Materials (DMR-0819860).

\newpage

\begin{figure}[h]
\caption{{\em Momentum-space resolved polariton distribution as power is increased}. The polaritons are created by an incoherent pump at in a spot of diameter 12 $\mu$m on the sample with $\delta = - 3.2$ meV (the polaritons are mostly photon-like). The white parabola gives the energy dispersion of the polaritons at the point of creation in the low-density limit. a) 1 mW pump power.  The smearing of the distribution toward the left corresponds to the momentum gained by the particles as they accelerate downhill in the gradient of energy created by the variation of cavity thickness. b) The same excitation conditions but at pump power of 34 mW. The distribution has become mono-energetic and is shifted to the blue, as discussed in the text. c) The same excitation conditions but at pump power of 44 mW.}
\label{fig.kspace}
\end{figure}

\begin{figure}[h]
\caption{{\em Real-space images of the polariton motion}, for the same conditions as the data of Fig.~\protect\ref{fig.kspace}, but restricted to $k_{\|}\simeq 0$ ($\pm 2^\circ$ angle of acceptance). a) The hot carrier luminescence created by the laser excitation spot, in the low-density limit (1 mW pump laser in a spot of radius 12 $\mu$m). This shows the spatial extent of the excitation region where the polaritons are generated and the exciton cloud which remains at the point of creation. b) Real-space image of the polariton emission at 1 mW pump power. c) 34 mW. d) 44 mW.  The white dashed lines show the calculated potential energy profile felt by the polaritons, which is the sum of the slope due to the cavity width gradient and the localized exciton cloud at the laser excitation spot.}
\label{fig.realspace}
\end{figure}

\begin{figure}[h]
\caption{
{\em Density dependence in the three density regimes}. a) Intensity of the photon emission peak at $k_{\|}=0$ (emission normal to the surface of the sample) as a function of excitation power, for conditions very similar to those of Figs.~\protect\ref{fig.kspace} and \protect\ref{fig.realspace}. (Laser spot size of 8 $\mu$m diameter, excitation spot at the point on the sample with lower polariton energy = 1.5954 eV.) Thin dashed line: linear dependence  (slope $=1$ on the log-log plot).  b) The fraction of the total luminescence contained in a spectral range of 0.5 meV centered on the wavelength of the peak emission intensity, integrated over all momenta, for the same conditions as those of (a).   c) The full width at half maximum of the spectral peak, from the same spectra as used for (a). The spectral resolution was 0.1 meV. The vertical dashed line at low density in each case indicates the point of deviation from linearity, which is also the point at which the coherent mono-energetic flow seen in Fig.~\protect\ref{fig.kspace}(b) begins to appear; the vertical dashed line at high density indicates the threshold pump power for the compact condensation shown in Fig.~\protect\ref{fig.realspace}(d).}
\label{fig.thres}
\end{figure}

\begin{figure}[h]
\caption{Two-dimensional far-field $k$-space images of the emission at different times, with 5 ms time averaging, from the trapped polariton state seen in Figs.~\ref{fig.kspace}(c) and \ref{fig.realspace}(d). The horizontal direction is in the direction of the cavity energy gradient, with downward to lower energy corresponding to motion to the right, in this case.}
\label{fig.vort}
\end{figure}

\newpage
\setcounter{figure}{0}

\begin{figure}
\begin{center}
\includegraphics[width=0.65\textwidth]{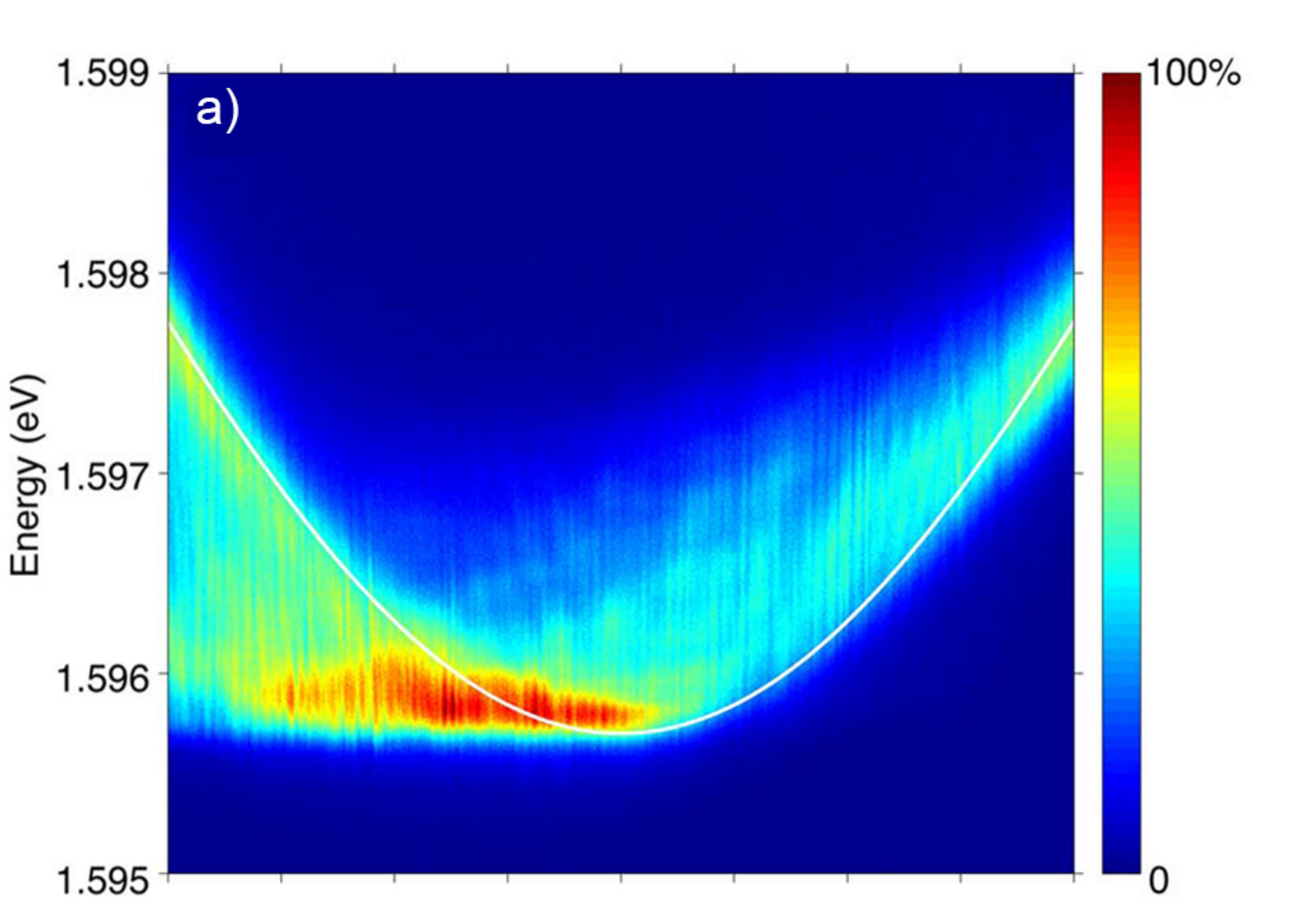}
\includegraphics[width=0.65\textwidth]{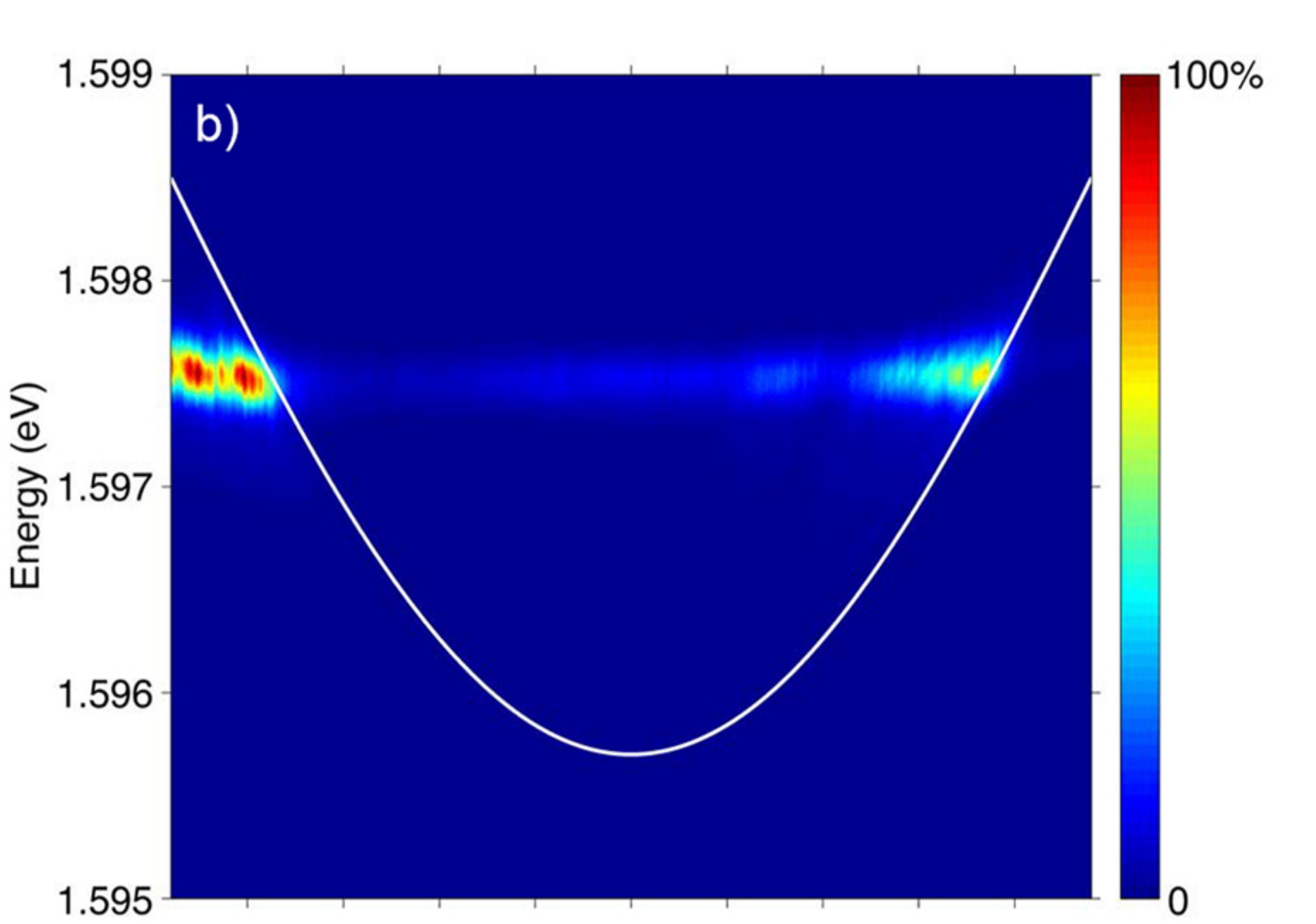}
\includegraphics[width=0.65\textwidth]{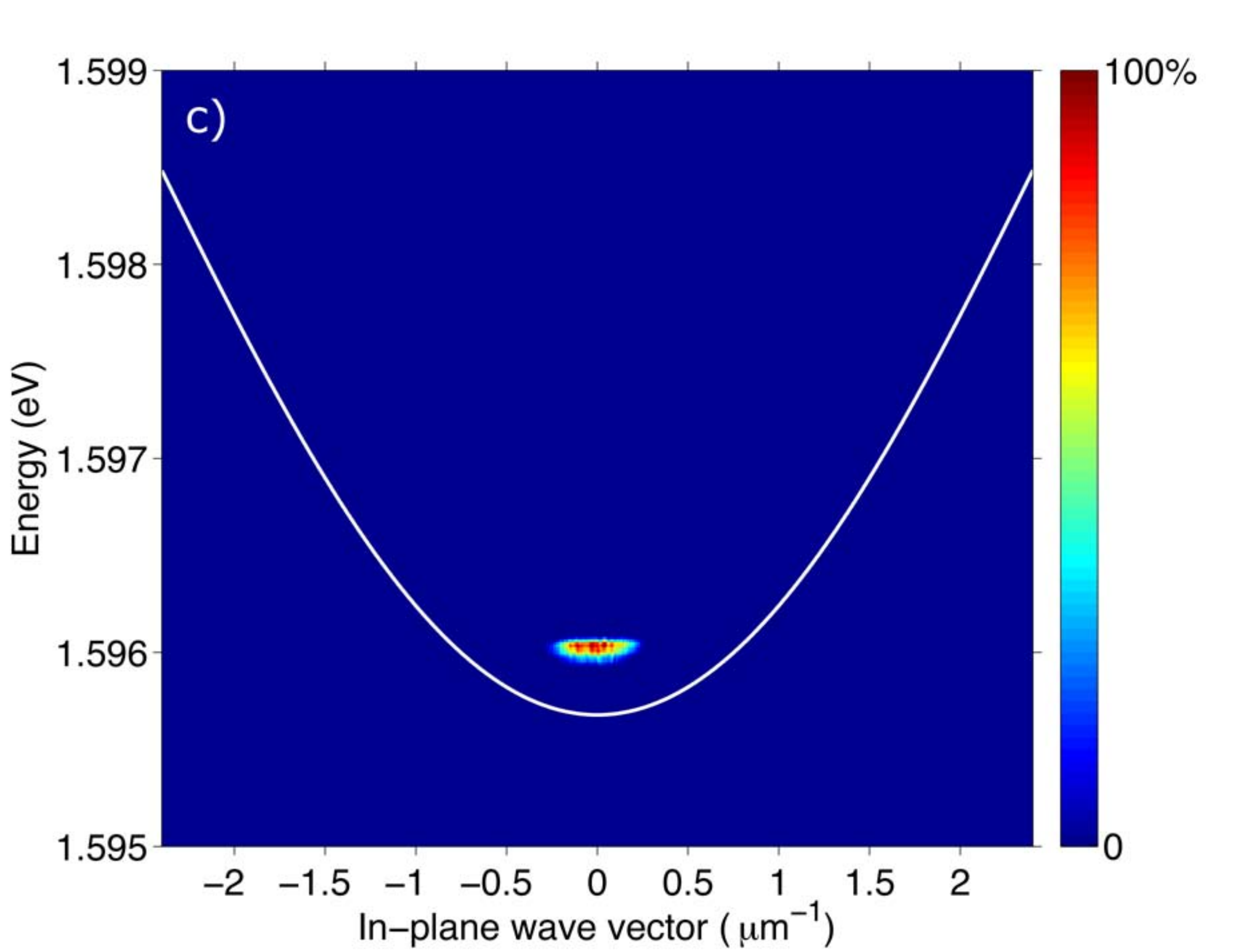}
\end{center}
\caption{ }
\end{figure}

\newpage

\begin{figure}
\hspace{.29cm}
\includegraphics[width=0.63\textwidth]{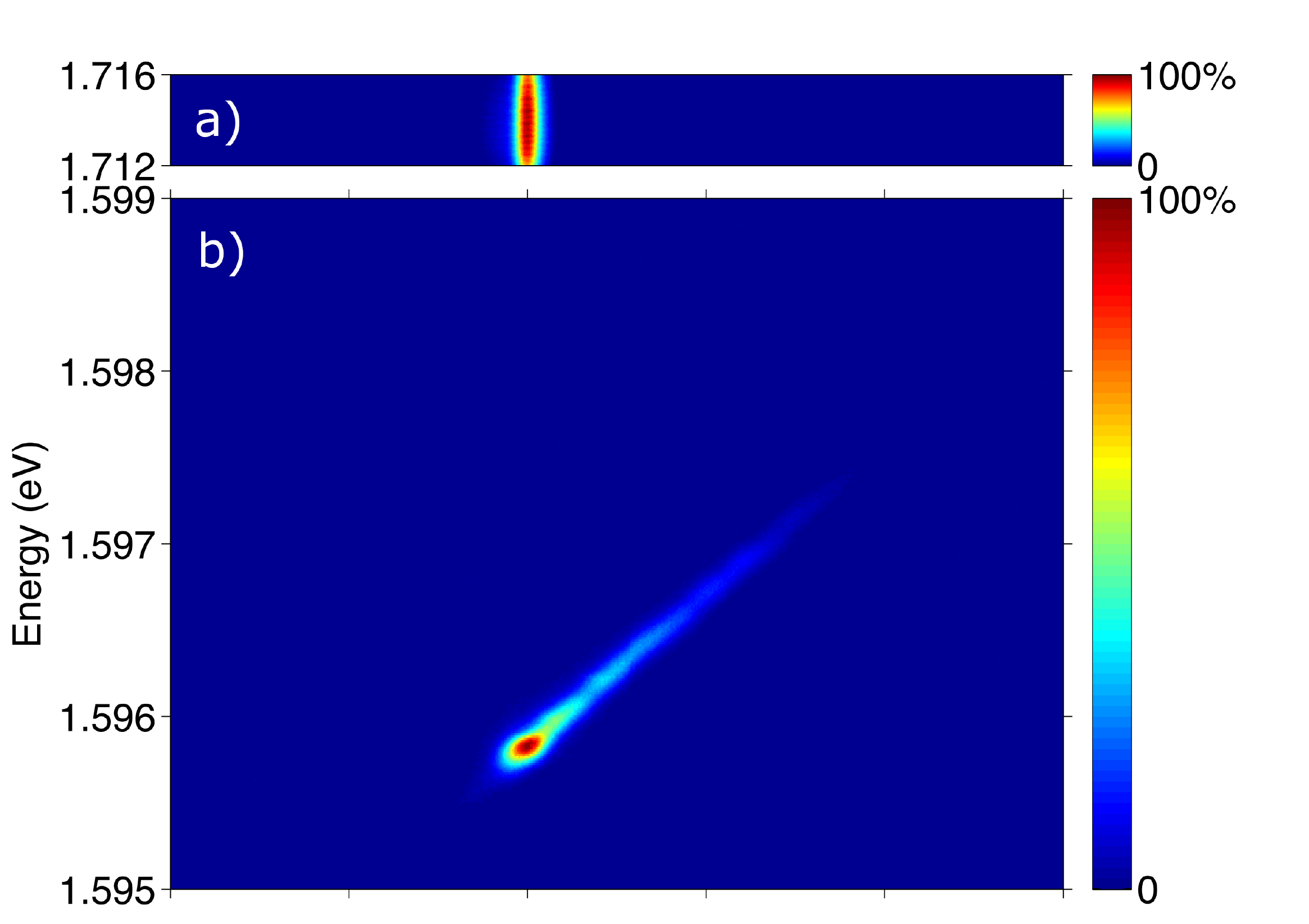}

\includegraphics[width=0.6\textwidth]{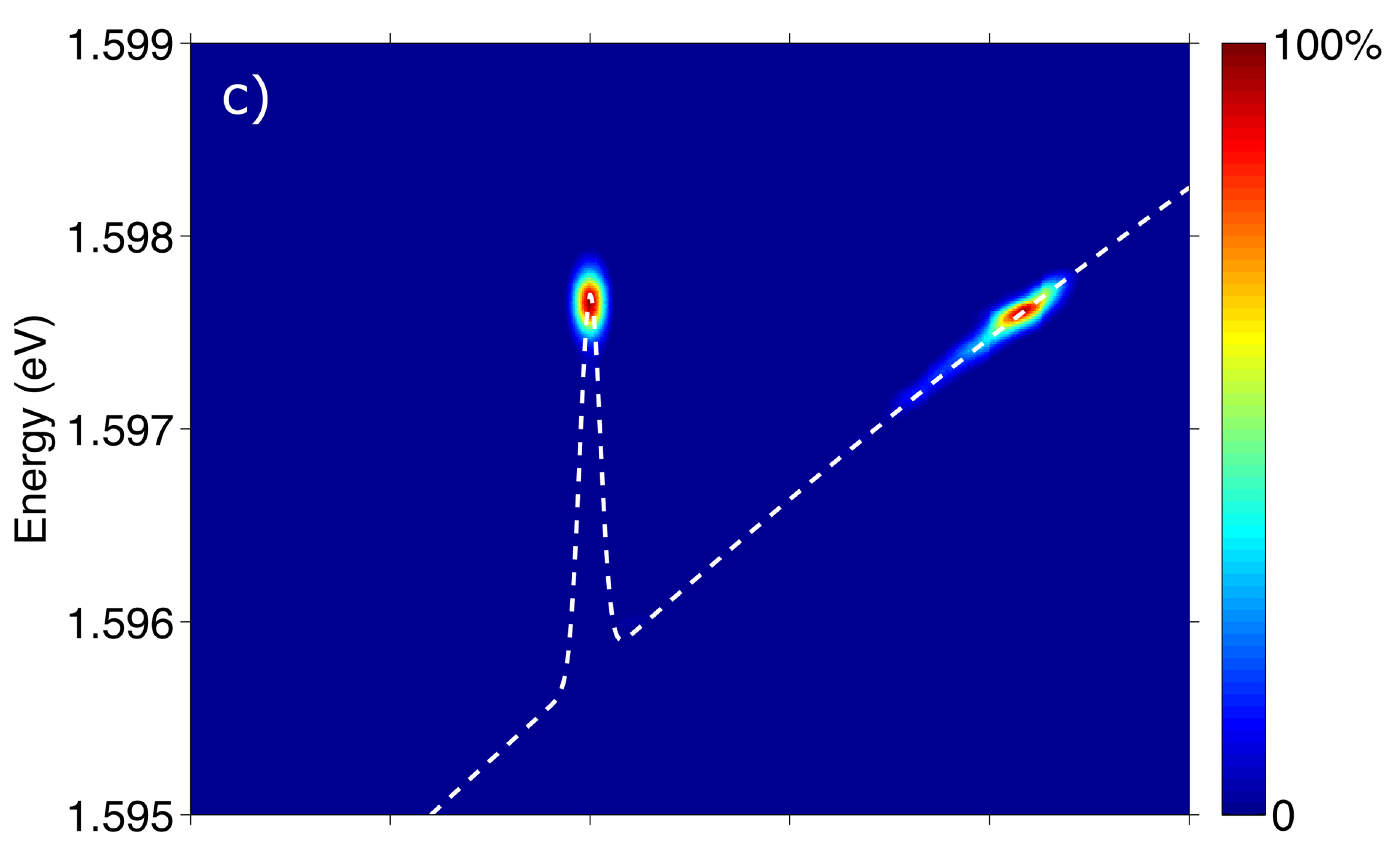}

\includegraphics[width=0.63\textwidth]{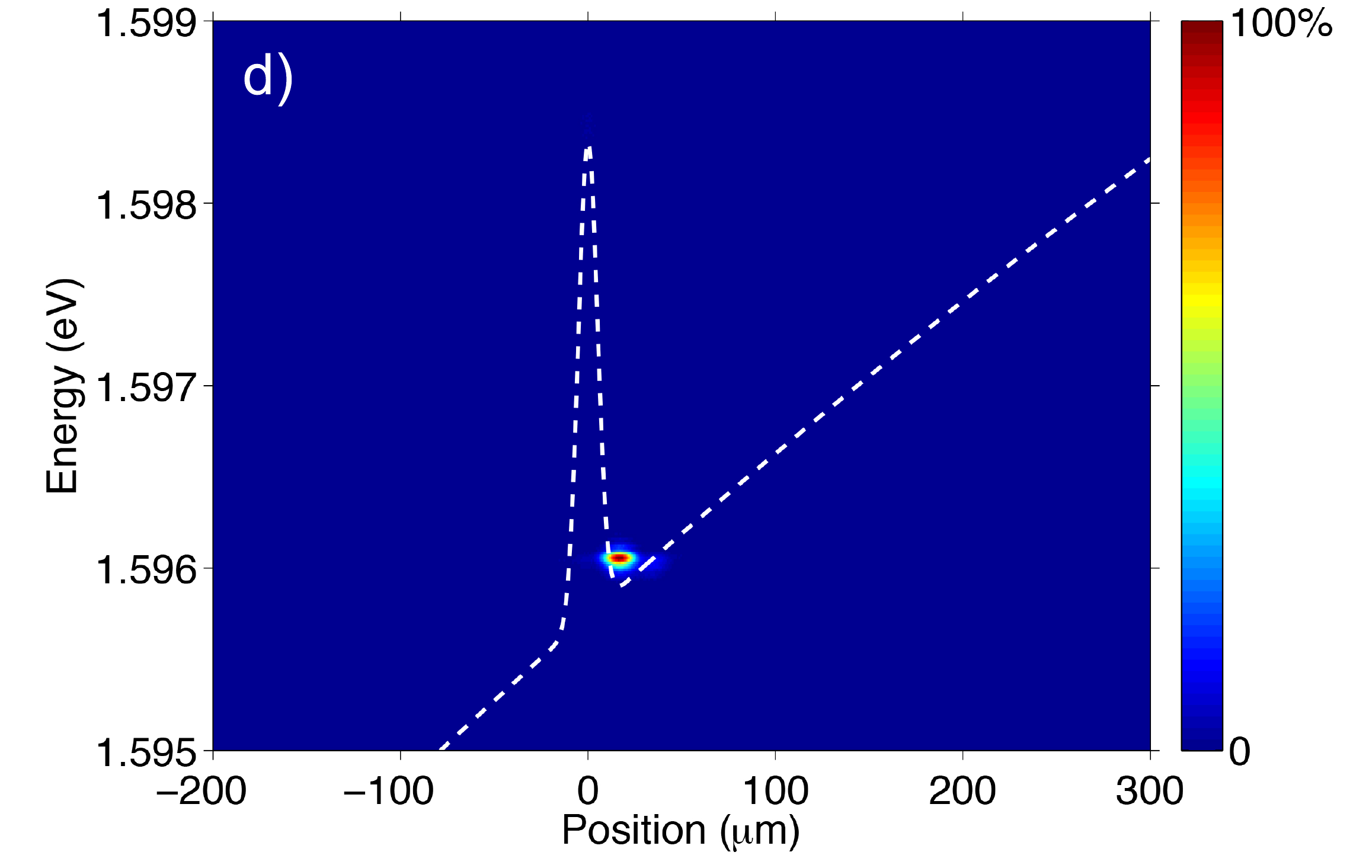}
\caption{ }
\end{figure}

\newpage

\begin{figure}
\begin{center}
\includegraphics[width=0.57\textwidth]{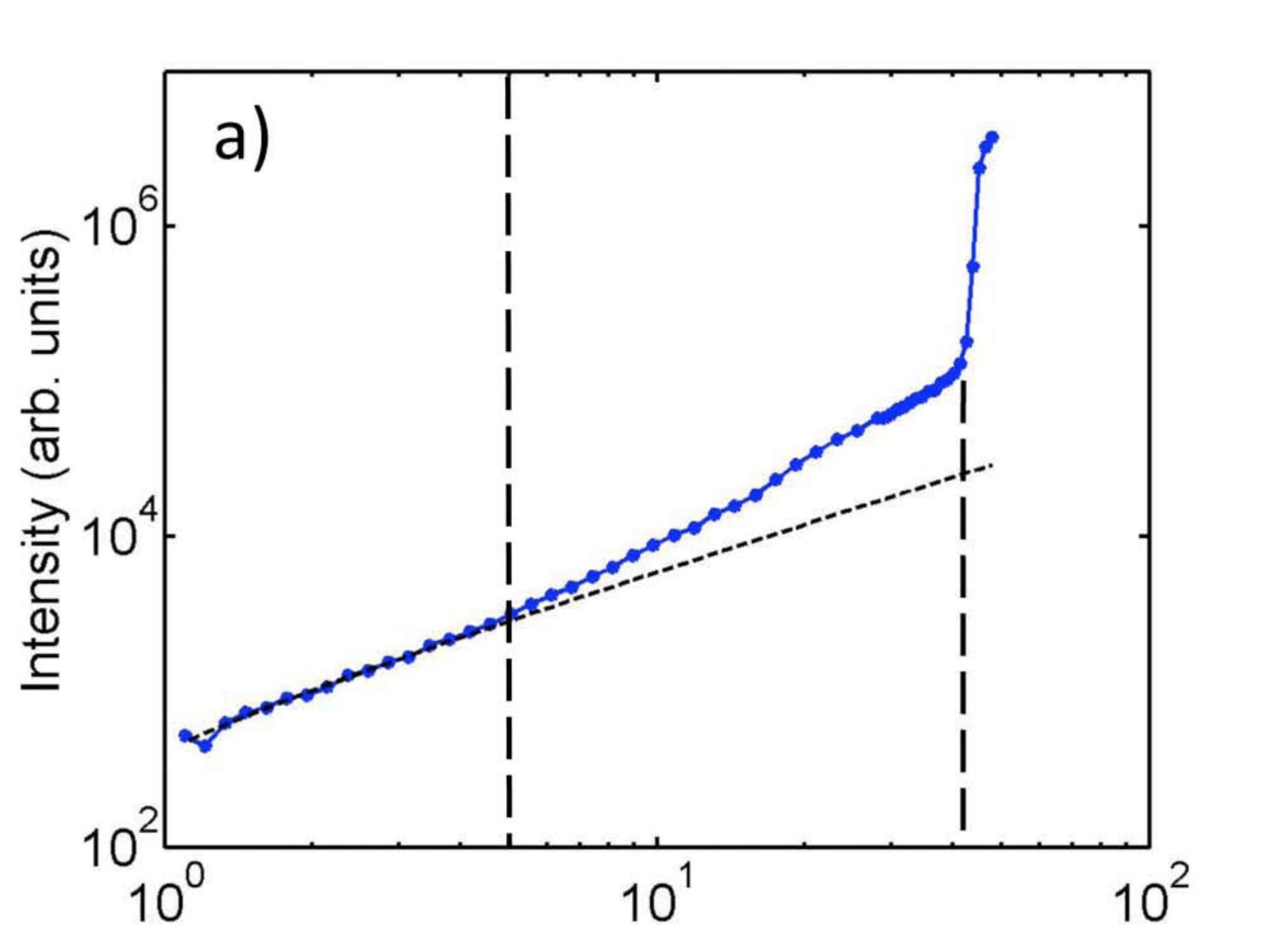}

\includegraphics[width=0.57\textwidth]{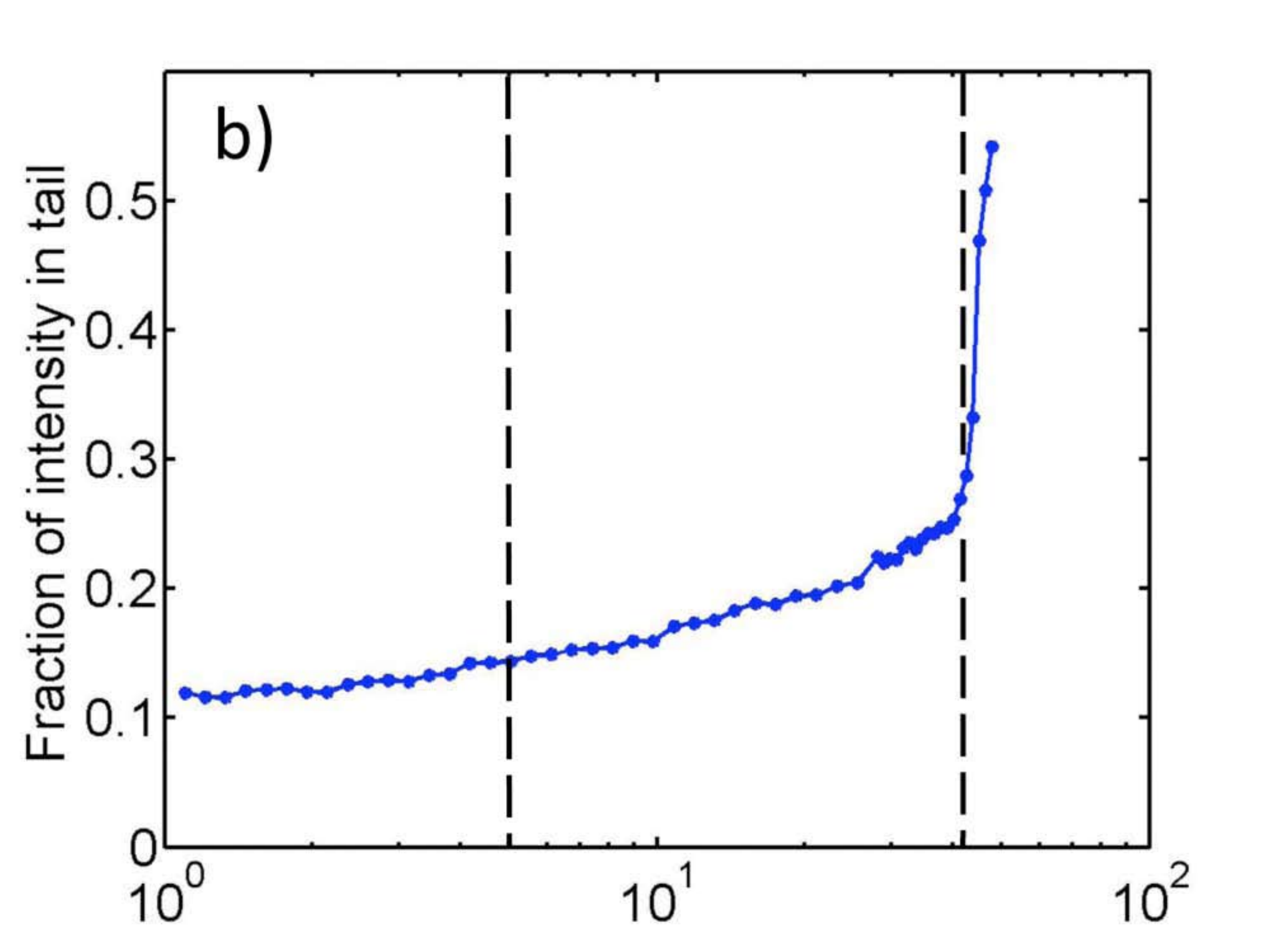}

\includegraphics[width=0.57\textwidth]{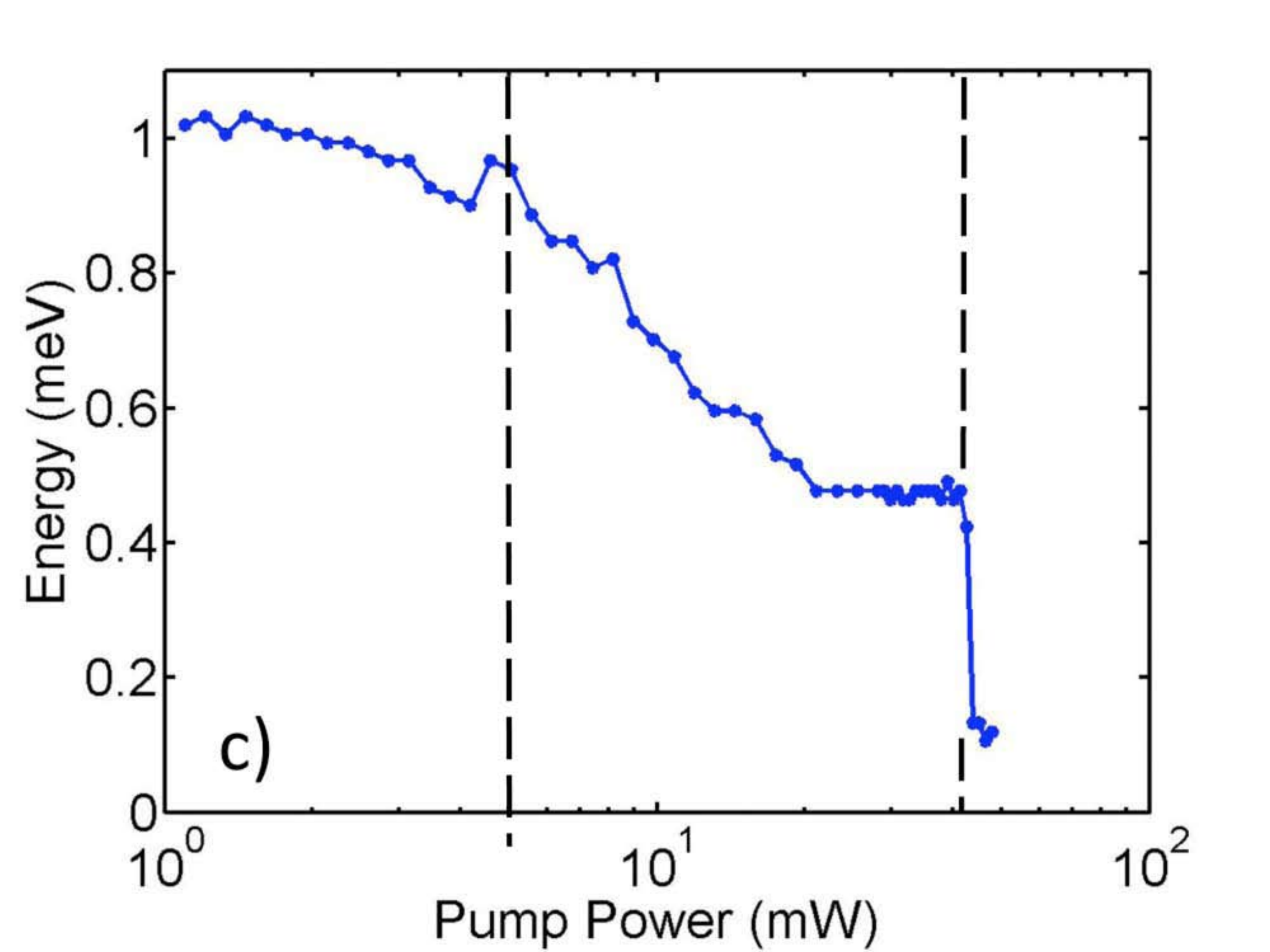}

\end{center}
\caption{ }
\end{figure}

\begin{figure}
\begin{center}
\includegraphics[width=0.66\textwidth]{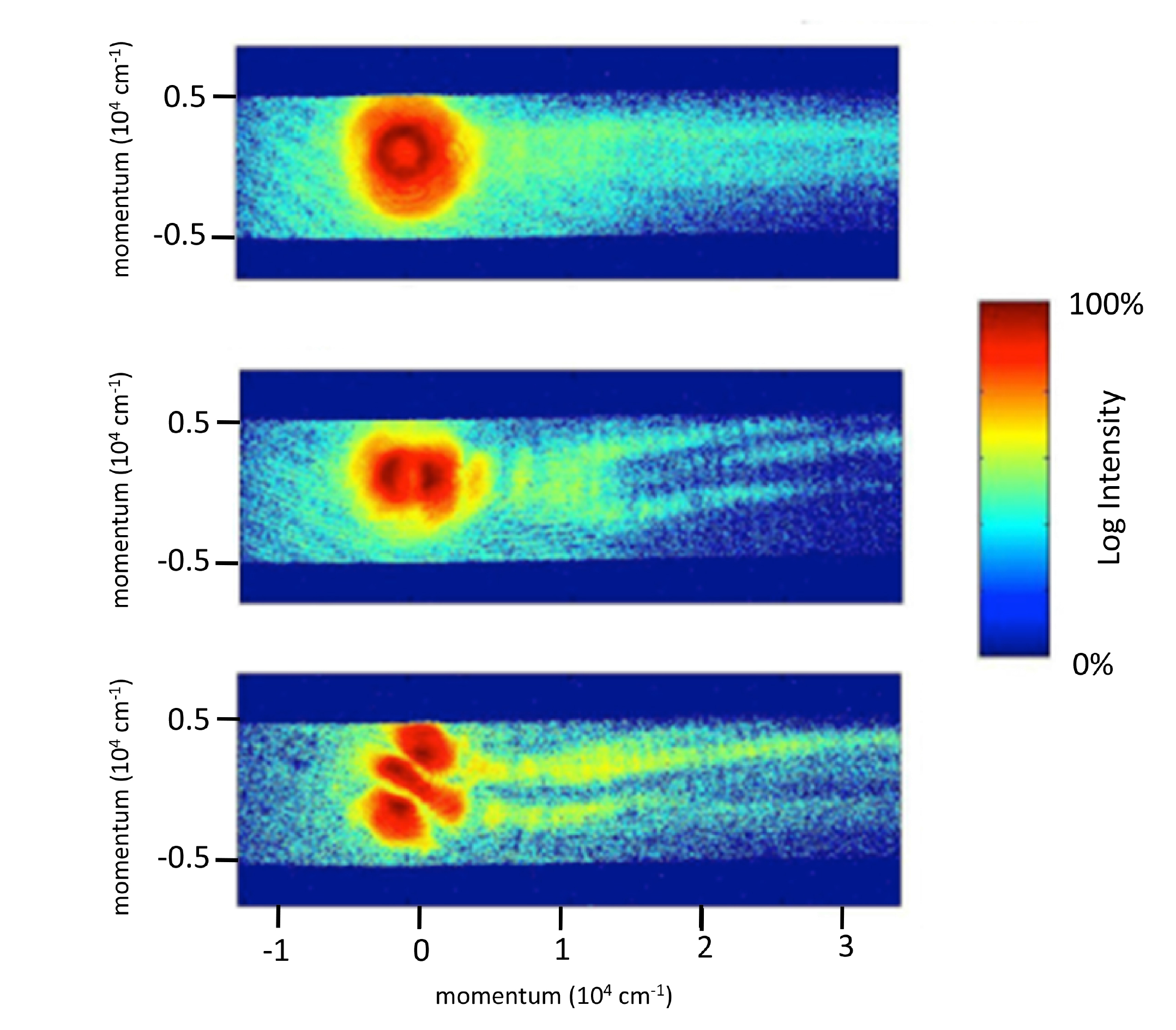}
\end{center}
\caption{ }
\end{figure}






\begin{thebibliography}{99}
\bibitem{wertz2} We are also able to select single spatial regions for $k$-space resolution using a pinhole in a secondary image plane as opposed to integrating over all space as in Fig.~\protect\ref{fig.kspace} here. Cf. Ref.~\protect\cite{bloch-wire}.
\end{thebibliography}
\end{document}